\documentclass[oneside,reqno,english]{amsart}
\usepackage[T1]{fontenc}
\usepackage[latin9]{inputenc}
\usepackage{geometry}
\usepackage{textcomp}
\usepackage{amsthm}
\makeatletter
\numberwithin{equation}{section}
\numberwithin{figure}{section}
\pdfoutput=1 
\usepackage{etoolbox}
\makeatletter
\patchcmd{\@maketitle}
  {\ifx\@empty\@dedicatory}
  {\ifx\@empty\@date \else {\vskip3ex \centering\footnotesize\@date\par\vskip1ex}\fi
   \ifx\@empty\@dedicatory}
  {}{}
\patchcmd{\@adminfootnotes}
  {\ifx\@empty\@date\else \@footnotetext{\@setdate}\fi}
  {}{}{}
\makeatother
\makeatother

\usepackage{babel}
\begin{document}
\title{On the distinction between beables and qualia}
\author{Adam Brownstein$^{*}$}
\thanks{$^{*}$Melbourne, Australia. ORCID: https://orcid.org/0009-0001-7814-4384\\
}
\date{February 11, 2025}
\begin{abstract}
\noindent The link between the de Broglie-Bohm interpretation and
the hard problem of consciousness is investigated. We highlight that
beables of the de Broglie-Bohm interpretation are not themselves qualia,
but are intimately related to qualia. It is then argued that qualia
are necessary for the de Broglie-Bohm interpretation, and conversely
the de Broglie-Bohm interpretation (or similar theories) are necessary
for qualia. Therefore, there is a reflexive two-way relationship between
the beables of the de Broglie-Bohm interpretation and qualia such
that they support the existence of each other. By using this reflexive
argument, the hard problem of consciousness can be addressed via an
indirect route.
\end{abstract}

\maketitle

\section{Introduction}

\noindent The hard problem of consciousness \cite{key-1} poses a
difficult question; and indeed a very important question central to
our understanding of the world. The hard problem of consciousness
suggests there is a difference between the physical object, and the
subjective first-person experience from the perspective of that object.
Subjective experience, which is known as qualia, is difficult to explain
from any material understanding of the particles or fields that underpin
the physical dynamical picture. 

For instance, if we view the brain as a biological network of neurons,
this would explain from a physical perspective the functions of seeing,
hearing, feeling and so forth; but not actually how these physical
processes evoke sound, imagery, and feeling in our subjective experience.

This is a peculiar fact assuming a reductionist view of the world.
For instance, an ant colony is known to display hallmarks of intelligent
behavior at the collective level. Yet presumably each ant is only
conscious of its own small world, and from the reductionist paradigm
it is clear that the colony itself does not have a subjective experience
or qualia. And similarly for a colony of bees, which are able to orchestrate
complex foraging behavior through chemical and visual messaging systems;
we don't suspect the hive of bees has a subjective first-person experience
of `what it's like to be a hive', even if there is intelligent behaviour
at the aggregate level. 

It follows that if neurons, like ants and bees, are divided from one
another (in this case by synapses and membranes), there should be
no collective consciousness of the network, and no subjective first-person
experience or qualia. Yet when it comes to the collection of neurons
in the brain, this is precisely what occurs. The problem is a paradox,
and progress is unfortunately inherently limited by the non-material
nature of the phenomena. However, in this paper we suggest that there
are hints toward an novel perspective on the problem in the relationship
between the de Broglie-Bohm interpretation and the existence of qualia.
There are two missing links which have brought us to this perspective. 

\section{The hard problem of consciousness}

\subsection*{First missing link: The de Broglie-Bohm interpretation does not solve
the hard problem of consciousness}

The de Broglie-Bohm interpretation, despite its ability to address
the measurement problem of quantum mechanics, does not solve the hard
problem of consciousness. At a surface level the theory may appear
to do so, because it is explicitly dualistic. There is matter as represented
by the wavefunction, and mental phenomena as represented by the particle
configuration or beables. Therefore it appears this was the problem
all along; that there is an explicit split between mind and body,
similar to Cartesian dualism. And because the particle configuration
is holistic; for example it has configuration space dynamics due to
entanglement effects; it appears exactly as what is required to solve
the hard problem of consciousness, because the system is not reducible
to parts, thus avoiding the issue of reductionism. 

However, there are several problems in this viewpoint when considered
more deeply. Firstly, the de Broglie-Bohm interpretation really is
quite like Newtonian mechanics in terms of its ontological entities,
despite these entities having the more complex quantum behaviour.
For example, it still represents a cat as a collection of particles
which traces out the pattern of a cat. And it still represents the
brain of the cat as a distribution of particles in the pattern of
neurons firing. None of this explains how the particle distribution
of the cat is able to ``see'' from the subjective first-person perspective
of `what it's like to be a cat'. The hard problem of consciousness
is how to get from particle distribution to qualia, not how to get
the particle distribution to conform to the quantum dynamics. Even
a brain depending on quantum mechanical effects would have this problem. 

You could alternatively hypothesize that the beables of the de Broglie-Bohm
interpretation are made of fields instead of particles. But neither
do fields explain the existence of qualia, they just make the issue
more confused by being irreducible to parts. Yet the hard problem
of consciousness is essentially a no-go theorem (e.g. like Bell's
theorem), which says that there is no possible material explanation
whatsoever. Indeed, the hard problem of consciousness has no obvious
solution; not particles, not fields, not the de Broglie-Bohm interpretation,
not extra-dimensions, not the Many-Worlds interpretation. All of these
things are physical explanations for the mechanics, but do not explain
the existence of a subjective first-person experience of `what it's
like to be a thing?'\footnote{For instance Nagel's `What it's like to be a bat?' \cite{key-3}. }.

The experience of `what it's like to be a thing' is something separate
from the physical matter, as demonstrated by the philosophical zombie
argument. It is conceivable to have a world of physical matter undergoing
mechanical time-evolution, but not experiencing the property of `what
it's like to be' that physical matter. This is also starkly demonstrated
by the reductionist arguments for the hard problem of consciousness
\footnote{For example the mill argument proposed by Leibniz \cite{key-2}.}.
However, reductionism only highlights the ontological issue which
extends deeper, to the non-reducible case. 

Furthermore, there can be no physical explanation for qualia, because
this merely shifts the goalposts. Any time a new physical structure;
for instance a particle, a field, or a hidden variable; is invented
to explain qualia, the hard problem of consciousness is now applicable
to that new physical structure instead of the original one. This is
precisely what has occurred in the de Broglie-Bohm interpretation.
It was designed to explain the measurement problem and the Born-rule
probabilities, and therefore in essence, explain qualia. But in doing
so, the hard problem of consciousness has shifted onto the beables
of the de Broglie-Bohm interpretation.

If the de Broglie-Bohm interpretation does not solve the hard problem
of consciousness, then what does it do? What is the utility in solving
the measurement problem of quantum mechanics without anyone (i.e.
qualia) to observe the measurements in the first place? In our view,
either one retains the false assumption that the de Broglie-Bohm interpretation
solves the hard problem of consciousness, or one acknowledges that
the motivations for the de Broglie-Bohm interpretation are actually
something deeper. 

This is the first missing link, as the idea that the de Broglie-Bohm
interpretation might not solve the hard problem of consciousness opens
up the possibility that a world could exist where the de Broglie-Bohm
interpretation were true but qualia were absent. This sets up the
conditions for the forward direction for a reciprocal relationship
between beables and qualia. The second missing link addresses the
reverse direction of the argument. 

\subsection*{Second missing link: The problem of probabilities in the Many-Worlds
interpretation}

Why is the de Broglie-Bohm interpretation necessary in the first place?
The reason is the problem of probabilities in the Many-Worlds interpretation.
All experimental evidence points toward the fact we live in a world
governed by Born-rule probability outcomes. However, the amplitudes
of the wavefunction are not probabilities, they are the square-root
of probability and complex-valued. Therefore, there is an explanatory
gap in moving from the complex-valued, square-root world of wavefunction
amplitudes to the real-world of our experiences. To bridge the explanatory
gap, there needs to be way to link up events and provide a continuity
of experience that satisfies the Born-rule probability outcomes across
time, and this is precisely what the de Broglie-Bohm interpretation
achieves. 

The de Broglie-Bohm interpretation is very similar to the Many-Worlds
interpretation. They are both no-collapse interpretations of the wavefunction.
However the de Broglie-Bohm interpretation takes the wavefunction,
and adds an additional mechanism on top i.e. the particle configuration
or beables. The particle configuration is able to tell a coherent
story about the time evolution that matches the Born-rule probabilities.
Now if qualia exist, then they require a mechanism. For instance,
qualia could be attached to the branches of the wavefunction, or alternatively
to the de Broglie-Bohm particle configuration. If qualia are attached
to the branches of the wavefunction, it immediately runs into the
problem of probabilities.

When the qualia are attached to the branches of the wavefunction,
it almost gives the right answer. The decoherent branches of the wavefunction
contain representations of our brain-states, so if qualia are attached
to these branches, it is evident how this might produce the phenomenology
of our brain states, at least initially. However, it does not produce
a consistent picture across time. 

While the qualia might be attached to a typical world such as ours
initially, after a period of time-evolution, they would rapidly depart
from the condition of Born-rule probabilities. This is related to
the naive counting argument against the Many-Worlds interpretation.
The naive counting argument suggests that if all worlds simultaneously
exist, then the only meaningful probability measure is to treat each
branch of the wavefunction equally, ignoring the wavefunction amplitudes.
It is essentially suggesting that qualia are attached to the branches
of the wavefunction with uniform probability. This is what is at the
heart of the problem of probabilities, and introducing the language
of qualia to the Many-Worlds interpretation makes the problem of probabilities
starkly evident. 

For a cartoonish example, the qualia could enter a branch of the wavefunction
where a dice is rolled 30 times and it comes up with the same number
each time. These non-typical branches are commonplace in the Many-Worlds
interpretation, as the branching process is unconstrained by probability
information, which is carried as a meaningless complex-valued amplitudes.
The complex-valued amplitudes do not directly effect the branching
process, which is governed by the unitary, linear time evolution of
the quantum state. Therefore, there is a statistical dispersion of
branches within the wavefunction into having a growing multiplicity
of non-Born rule worlds. Only when taking the amplitude squared of
the wavefunction do these complex-valued amplitudes acquire a meaning. 

It is also not often discussed in the context of the Many-Worlds interpretation,
but the majority of wavefunction branches are unstable worlds that
contains off-shell behaviour. For instance, if qualia were attached
to a typical wavefunction branch, particles would jump out of the
electron orbitals frequently, and matter would decay. Take any initial
branch of the wavefunction, and the most likely outcome; treating
all branches equal irrespective of their amplitudes; is a world that
quickly disintegrates. Therefore, assigning qualia to the wavefunction
directly is untenable, unless there is a way to guide the qualia toward
the right branches as in the de Broglie-Bohm interpretation.

Consequently, the problem of probabilities in the Many-Worlds interpretation
provides the reverse direction of argument for a reciprocal relationship
between beables and qualia. Beables are required in addition to the
wavefunction, so that qualia have something meaningful to attach to.
Therefore, the de Broglie-Bohm interpretation (or similar theories)
are necessary for qualia. 

\subsection*{Are beables and qualia different things? }

Before continuing, we wish to make a digression and to discuss the
nature of beables and qualia further, for perhaps the beables of the
de Broglie-Bohm interpretation have been incorrectly interpreted as
something material, instead of non-material on the same footing as
qualia (and perhaps as the qualia themselves). While viewing the beables
as non-material is possible, the downside of this viewpoint is that
it risks placing a too great philosophical burden upon them. In particular,
the beables have a well-defined mathematical description of their
dynamics, and therefore a material description is evidently possible.
A realist view of nature might ideally explain as many elements of
reality as material physical objects as possible. Furthermore, making
the change in ontology would have little practical consequence if
non-material qualia are attached to the beables in any case. 

\subsection*{Do qualia exist? }

The existence of qualia is evident by an anthropic argument. If the
universe contained no qualia, there would be no conscious entities
present to acknowledge its existence in the first place. The reason
why we perceive ourselves to be here is precisely because we have
qualia. If we did not have qualia, we would not perceive ourselves
to be here. A hypothetical sterile universe, where everybody is a
philosophical zombie, is untenable as it contradicts basic human experience
of the world. We strongly believe that qualia is simply a brute fact
that is proven by the empirical observation of one's own direct experience.
This brute fact is important in the conclusions on the reciprocal
relationship between beables and qualia. 

\subsection*{Reification and the properties of qualia}

Whatever qualia is (e.g. a substance or property), it appears to have
a continuity of experience to it; an identity across time; and this
continuous experience matches the Born-rule probabilities not only
instantaneously but in the sense of time-evolution. This might be
because the phenomenon of qualia is attached in a metaphysical sense
to real physical objects which have continuous dynamics across time. 

Our suggested interpretation of qualia is that it is a metaphysical
reification (i.e. ``making real'') of a material substance, where
the material substance is given the property of `what it's like to
be' that substance, a necessary property for something to be considered
present in the physical world. For how can any physical object exist
in the world without the property of `what it's like to be' that object?
It is necessary to assign the property of `what it's like to be' an
object in order for it not to simply be a mathematical description,
but to physically exist in the world. The curious thing is not that
this step of metaphysical reification is required, it is the hard
problem of consciousness, namely why `what it's like to be' a physical
brain is conscious rather than non-conscious. In other words, the
reification step is evident, however the outcome of the reification
is quite peculiar. 

We envisage that the qualia inherit the properties of whatever they
attach to. Therefore, the nature of the game regarding qualia is to
identify the right physical structure to assign qualia to; i.e. a
physical structure which has a continuous identity across time, and
satisfies the Born-rule probabilities. This role is adequately filled
by the beables of the de Broglie-Bohm interpretation, but not by the
branches of the wavefunction. The phenomena of qualia may not necessarily
require any more justification than that; it seems to be a non-material
phenomena that is completely structureless, and must be attached to
material matter to give it physical meaning. 

The story-like nature of the qualia is evident in our experiences
of the world, and this is likely due to the material substance that
the property of qualia is attached to. The continuity of experience
of qualia would rule out Bohmian-like flash ontologies for the Born
rule probabilities. The qualia we experience are not a dust-like series
of random flashes that momentarily satisfy the Born-rule probability
distribution but lack the continuity across time. Nor is is sufficient
to attach the qualia to objects or fields that have continuous identity
across time, but do not satisfy the Born-rule probabilities across
time. This is the problem of attaching the qualia to the branches
of the wavefunction in the Many-Worlds interpretation. It gets the
continuity of identity correct, but not the continuity of satisfying
the Born-rule probabilities across the time-evolution. 

In our view, qualia may be considered akin to a platonic idea, which
exists independently of material substance. For example, we know that
physical matter follows equations of motion. These equations are only
valid because the mathematics of these equations has a truth value
in the platonic world of conceptual ideas. Therefore, the platonic
world provides a reification to the material side of physics. A similar
situation occurs with qualia, however the reification is more about
what it means for something to exist in the world in a tangible sense.
This reification is evidently necessary, because by the anthropic
argument, if we weren't conscious, then we wouldn't perceive ourselves
to be here to ponder the question. Therefore, the reification is a
brute fact of conscious existence. 

\section{A reciprocal relationship between beables and qualia}

\subsection*{Forward argument: Qualia are necessary for the de Broglie-Bohm interpretation }

Imagine that qualia did not exist. Then the de Broglie-Bohm interpretation
would be a theory of a de Broglie-Bohm particle distribution being
guided by the wavefunction as usual. However, the de Broglie-Bohm
particle distribution would become redundant, because the information
it contains is the same as the information contained in the wavefunction.
It would be an epiphenomenon which serves no purpose. It is only by
the existence of qualia and the experience of Born-rule probabilities
that there is a necessity to explain these Born-rule probabilities
in the first place. Therefore qualia are necessary for the de Broglie-Bohm
interpretation. 

This argument against redundancy is an appeal to something akin to
Leibniz's principle of sufficient reason. We suggest that the physical
construction of the universe should not have purposeless redundancy,
such as a de Broglie-Bohm particle distribution that lacks qualia,
in addition to the wavefunction. If qualia did not exist, then a Many-Worlds
ontology might be considered preferable. However, as the case stands,
qualia evidently do exist and therefore the de Broglie-Bohm interpretation
is viable. 

\subsection*{Reverse argument: The de Broglie-Bohm interpretation (or similar
theories) is necessary for qualia}

If we take qualia as a given, it is necessary to have a physical substance
to attach the qualia to. As discussed in previous sections, due to
the problem of probabilities of the Many-Worlds interpretation, it
is insufficient to attach the property of qualia directly to the wavefunction.
Instead, the qualia must be attached to a physical entity that a)
has a consistent identity across time and b) reproduces the Born-rule
probabilities across time. The beables of the de Broglie-Bohm interpretation
are entities which are designed to satisfy a) and b). Therefore the
de Broglie-Bohm interpretation (or similar theories which satisfy
these properties) are necessary for qualia. 

\subsection*{Inference 1: Beables and qualia share a reflexive two-way relationship }

We have made the claim that the de Broglie-Bohm interpretation requires
qualia, and separately that qualia require the de Broglie-Bohm interpretation.
What does this reflexive two-way relationship entail? Firstly, perhaps
these two things, of qualia and the beables of the de Broglie-Bohm
interpretation, are linked in a special way that has not yet been
understood. It is difficult to describe this connection without resorting
to a material explanation of qualia. Therefore, a more fruitful approach
may be given via pure logical reasoning. 

Taking a logical point of view on the construction of the universe,
perhaps beables and qualia entail each other into existence. Firstly,
why does anything exist? Because the wavefunction exists, which describes
a dynamics fully compatible with spacetime and relativity. Then, why
do qualia exist? Because we have the de Broglie-Bohm interpretation.
Then, why does the de Broglie-Bohm interpretation exist? Because we
have qualia. The argument is circular, but perhaps this is a feature
rather than a deficiency; because any linear argument must terminate
in the brute fact of the anthropic principle and thus will fail to
provide an explanation for why. 

\subsection*{Inference 2: Inverting the hard problem of consciousness }

It is actually difficult to say which is more important for the anthropic
argument. Perhaps the universe was constructed such that qualia were
true, and the only way to make it viable was to add beables. This
seems like the most obvious chain of logic. Yet we can take the opposite
chain of logic too. Perhaps the universe was constructed such that
the de Broglie-Bohm interpretation were true, and the only way to
make it viable was to add qualia. In this way, the the hard problem
of consciousness can be understood from a new perspective, as the
problem is inverted from qualia onto the beables of de Broglie-Bohm
interpretation. 

We shouldn't necessarily conclude that qualia is the most important
feature for the anthropic argument just because it is the one most
immediately available to our direct experience. If the qualia are
necessitated by something else, then that something else can be placed
at the center of the anthropic argument. Placing qualia at the center
of the anthropic argument is a nonmaterial-centric view, with material
justifications for existence, while placing beables at the center
of the argument is a material-centric view with non-material justifications
for existence. The second viewpoint may be more aligned with the common
understanding of the relationship between physics in the material
sense, and mathematics in the non-material platonic sense. The physics
typically is the central brute fact amenable to direct empirical observation,
while the mathematics provides the theoretical justification and logical
framework for its existence. 

Because qualia has no possible explanation in terms of material substance,
it might not be the best starting point for constructing a logical
explanation for its existence. It might be preferable to invert the
problem so that it is centered on something material e.g. the beables.
For example, it would be elegant if the de Broglie-Bohm interpretation
were true, therefore taking a material-centric view, but the only
way to make it work was to add the mysterious phenomena of qualia;
just an added extra that has no explanation and no material properties,
a benign addition to the theory like a zeroth-order platonic ideal.
The governing principle seems to be Leibniz's principle of sufficient
reason, working in conjunction with the anthropic principle. Objects
should not have a redundancy in their physical description unless
it is strictly necessary, for example due to supporting the existence
of qualia. 

\section{The metaphysical status of the time coordinate in physics}

\subsection{Is the time-coordinate on the same footing as qualia?}

An important point regarding the time-coordinate in physics can also
be made. There are three prevalent views of the time coordinate. Firstly,
it can be viewed as an abstract parameter, in the mathematical sense
as an external dimension that allows the dynamics to be visualized.
This is a common viewpoint, taking the pure materialist perspective
of nature. For instance, in classical mechanics, matter exists in
three-dimensional space and the time coordinate is an additional parameter
used to make the equations work. Therefore, the most basic interpretation
is to regard time as an abstract parameter or concept that exists
in the platonic realm. 

Secondly, the time coordinate is sometimes viewed as part of the physical
world, having the same status as the spatial dimensions. This often
implies a static block universe which is four-dimensional. This idea
is also compatible with a materialist view of the world. Thirdly,
it can be assumed not to exist, and there is only one time which is
the now. This viewpoint does overlap as variant of the first, however
we have included it for completeness as a separate perspective. 

Nevertheless, perhaps the real status of the time-coordinate is not
that it is part of the platonic or material worlds, but instead it
may be part of the non-material world, on the same footing as qualia.
The distinction between the material (i.e. physical stuff), non-material
(i.e. experiential phenomena) and platonic (i.e. mathematical and
logical concepts) needs to be made. It does not make sense to have
time as part of the platonic world, as this is reserved for abstract
concepts such as the truth value of mathematical equations or geometric
forms, whereas the time parameter is something quite real, just distinct
from the spatial coordinates which are of a more physical nature. 

This is especially important in general relativity. We must distinguish
between the local time coordinate, which may be part of the physical
four-dimensional spacetime, and the proper time, which is an abstract
parameter similar to the Newtonian time in classical mechanics. Proper
time is often envisaged as part of the platonic world, as a mathematical
parameter. However, proper time might be more correctly considered
as part of the non-material world, having the same basic status as
qualia. 

Furthermore, proper time can be measured, or at least inferred by
calculating the relativistic arc-length. Therefore it may be possible
to measure an aspect of the non-material world, using material physical
instruments. Curiously, the measurement of proper time takes place
by comparing different local physical measurements to extract global
information, for instance the relativistic arc-length, and curvature
of the spacetime; which is a concept that exists outside the local
four-dimensional coordinate system, and is a feature of either the
platonic or non-material worlds. Note that curvature of the four-dimensional
local coordinates can only be visualised in the fifth dimension, and
this fifth dimension is either of a platonic or non-material (i.e.
of a qualia-like) nature. Further thought into the relation between
proper time and qualia may be required, and to the relation between
the non-material world and the presence of global, relational or curvature
information that can be extracted by comparing a collection of local
measurements. 

An interesting fact about viewing time as a non-material aspect of
nature, is that unlike qualia, it isn't about subjective experience.
Even though individuals experience time, matter exists within time
regardless of whether sentient life is able to observe it. Therefore,
this provides a challenge to the viewpoint that the non-material world
is purely about experience or qualia. There may be other relevant
non-material phenomena beyond qualia, for instance global curvature,
topology, entropy and relational meaning. We furthermore conjecture
that there is a conceptual error in envisaging the physical world
as embedded in an abstract platonic space (for instance viewing the
four-dimensional spacetime coordinates embedded in an abstract fifth
dimension). Instead, it may be logically preferable to view the physical,
material world as embedded in the physical, non-material world, not
in an abstract platonic space which is non-physical and conceptual. 

\section{Conclusions}

\noindent We have explored in this paper the relationship between
beables of the de Broglie-Bohm interpretation and qualia. It is suggested
that beables are not qualia, but are intimately linked to qualia in
a reciprocal relationship. In the forward direction of the argument,
we argue that qualia require something like beables to attach to,
so as to provide a continuous dynamical picture across time, and also
to satisfy the Born-rule probabilities across time. This viewpoint
is necessitated by the problem of probabilities in the Many-Worlds
interpretation, which indicates qualia cannot be attached directly
to the wavefunction. In the reverse direction of the argument, we
have used Leibniz's principle of sufficient reason to argue that beables
require qualia, so as to avoid redundancy in addition to the existence
of the wavefunction. 

Therefore in summary, beables require qualia for their existence,
and qualia require beables for their existence. This curious two-way
relationship may hold clues into the nature of the physical world,
which has both material and non-material aspects, as the hard problem
of consciousness has shown. Viewing the problem of consciousness in
terms of beables opens up new avenues for logical thought. In particular,
the reciprocal relationship can be used to invert the hard problem
of consciousness from qualia onto the beables.

The forward viewpoint is that the universe was constructed such that
qualia exist, which is taken to be a logical necessity by the anthropic
argument. Then beables were required to provide something for the
qualia to attach to, and the wavefunction was required to guide the
beables. The reverse viewpoint is that the universe was constructed
such that the beables and wavefunction exist, again motivated by the
anthropic argument. However, this means the beables would provide
redundant information in comparison to the wavefunction. Therefore,
by Leibniz's principle of sufficient reason, qualia were a necessary
addition to avoid redundancy of the beables. 

This may seem like a convoluted chain of logic, but it does explain
two things. Firstly, it provides an explanation of why we have the
non-material phenomena of qualia. The answer given is that qualia
are necessary to avoid redundancy of the beables in addition to the
wavefunction, due to Leibniz's principle of sufficient reason. So
the question of why qualia exist has an answer which is not simply
a brute fact rooted in the anthropic argument. The remaining question
is why does quantum mechanics have the structure that it does e.g.
why are beables and the wavefunction both necessary? The necessity
of the wavefunction may be due to the requirement of satisfying basic
physical principles such as special relativity and gauge transformations,
which require the complex-valued algebra of spin matrices. Then the
de Broglie-Bohm particles are required to generate the Born-rule classical
world from the basic substance of the wavefunction. 

Contrast this to the forward interpretation, where qualia are regarded
as necessary due to the anthropic argument; which is to say that if
qualia did not exist, then we would not be conscious of asking the
question. Although the anthropic argument forces the conclusion that
qualia are necessary in our world, it cannot explain why the world
was not non-sentient in the first place, only that it must have sentient
matter if it is to be observed. This is not a sufficient answer to
the question of why qualia and consciousness are required, because
the answer returns to the brute fact of the anthropic argument. Conversely,
by regarding qualia as necessary for the beables to exist, we do find
a possible means to escape the anthropic argument as means of explanation. 

Secondly, it places the material objects in their proper place at
the center of the argument. For example, if one considers the logical
framework for constructing a typical Newtonian dynamical picture,
one might imagine that there are particles which exist in the world
in ordinary three-dimensional space. Then a time coordinate must be
introduced to allow dynamical motion. This time coordinate is regarded
as an abstract parameter that is required to make the material picture
work. Analogously, using this inverted chain of logic on the relationship
between beables and qualia, the material world of beables and the
wavefunction are taken to be primary. However, to make the beables
work, the abstract phenomena of qualia must be added to the theory. 

The two situations are similar in their basic logical setup. Both
start with the material phenomena to be explained at the center of
the argument, and then an additional abstract parameter or phenomena
that allows the theory to work is added to the picture. The similarity
between the two scenarios may be even more evident, when we consider
that the time coordinate might be more correctly regarded as being
on the same footing as qualia as a physical, non-material aspect,
and not a part of the non-physical, platonic world of concepts and
mathematical truths. 

\section*{Statements \& declarations}

\subsection*{Statement of originality}

All work is original research and is the sole research contribution
of the author. 

\subsection*{Conflicts of interest: }

No conflicts of interest. 

\subsection*{Funding: }

No funding received. 

\subsection*{Copyright notice: }

Copyright{\small{} }{\footnotesize{}© }2024 Adam Brownstein under
the terms of arXiv.org perpetual, non-exclusive license 1.0.


\begin{thebibliography}{1}
\bibitem{key-1} Chalmers, D. J. (1995). Facing up to the problem
of consciousness. Journal of consciousness studies, 2(3), 200-219.

\bibitem{key-3} Nagel, T. (1980). What is it like to be a bat?. In
The language and thought series (pp. 159-168). Harvard University
Press.

\bibitem{key-2} Leibniz, G. W. (1989). The Monadology: 1714 (pp.
643-653). Springer Netherlands.
\end{thebibliography}
\end{document}